\documentclass[12pt,preprint]{emulateapj}

\shorttitle{NGC 3311}
\shortauthors{Wehner et al.}

\begin{document} 
\title{The Globular Cluster Systems Around NGC 3311 and NGC 3309
}

\author{Elizabeth M.~H.~Wehner}
\affil{Department of Physics \& Astronomy, McMaster University, Hamilton L8S 4M1, Canada}
\email{wehnere@physics.mcmaster.ca} 

\author{William E.~Harris} 
\affil{Department of Physics \& Astronomy, McMaster University, Hamilton L8S 4M1, Canada}
\email{harris@physics.mcmaster.ca} 

\author{Bradley C.~Whitmore}
\affil{Space Telescope Science Institute, 3700 San Martin Drive, Baltimore MD 21218}
\email{whitmore@stsci.edu}

\author{Barry Rothberg}
\affil{Space Telescope Science Institute, 3700 San Martin Drive, Baltimore MD 21218}
\email{rothberg@stsci.edu}

\author{Kristin A.~Woodley} 
\affil{Department of Physics \& Astronomy, McMaster University, Hamilton L8S 4M1, Canada}
\email{woodleka@mcmaster.ca}

\begin{abstract} 
We present extensive new photometry in $(g',i')$ of the large globular
cluster (GC) system around NGC 3311, the central cD galaxy in the Hydra cluster.
Our GMOS data cover a $5\farcm5$ field of view and reach a 
limiting magnitude $i' \simeq 26.0$, about 0.5 magnitude fainter
than the turnover point of the GC luminosity function.  
We find that NGC 3311 has a huge population of $\simeq 16,000$ GCs, 
closely similar to the prototypical ``high specific frequency''
Virgo giant M87.
The color-magnitude distribution shows that the metal-poor ``blue'' GC sequence
and the metal-richer ``red'' sequence are both present,
with nearly equal numbers of clusters.  Bimodal fits to the
color distributions confirm that the
blue sequence shows the same trend of progressively increasing
metallicity with GC mass that has previously been found in many other
large galaxies; the correlation we find corresponds to a scaling of
GC metallicity with mass of $Z \sim M^{0.6}$.
By contrast, the red sequence shows no change of mean metallicity
with mass, but it shows an upward extension to much higher
than normal luminosity into the UCD-like range, strengthening the
potential connections between massive GCs and UCDs.
The GC luminosity function, which we
measure down to the turnover point at $M_I \simeq -8.4$, also
has a normal form like those in other giant ellipticals.
Within the Hydra field, another giant elliptical NGC 3309
is sitting just $100''$ from the cD NGC 3311.  We use our data  
to solve simultaneously for the spatial structure and total GC
populations of both galaxies at once.  Their
specific frequencies are $S_N(N3311) = 12.5 \pm 1.5$ and
$S_N(N3309) = 0.6 \pm 0.4$. NGC 3311 is completely dominant
and entirely comparable with other cD-type systems such as M87 in Virgo.  
\end{abstract}


\section{Introduction} 
\label{intro}

One of the most distinctive features of the supergiant cD-type galaxies
found at the centers of rich environments is the very obvious presence of
huge populations of old-halo globular clusters (GCs), often numbering
above 10,000 GCs in a single system and spanning
radii upwards of 100 kpc \citep[e.g.][]{harris01}.
These ``high specific frequency'' (``high-$S_N$'') globular cluster systems (GCSs) provide
a platform for statistical studies of halo cluster properties that cannot be
carried out in any other environment.  Despite their rarity, they are 
therefore prime targets for both observations and interpretive modelling on the
formation and evolution of GCs and their host galaxies.

The GCSs in cD galaxies (most of which are also ``Brightest Cluster Galaxies''
or BCGs) have been used to identify a new correlation between
GC luminosity (or mass) and color (or metallicity).  While the cluster metallicity
distribution function (MDF) usually has a 
bimodal shape \citep[e.g.][among many others]{zepf93,larsen01,mieske06}, 
recent photometric work suggests that the 
more metal-poor GCs (those in the ``blue sequence'') have higher
heavy-element abundance at progressively higher luminosity.
This {\sl mass-metallicity relation} (MMR) was first discovered in eight giant
BCGs by \citet[][hereafter H06]{har06} and in the Virgo giants M87 and
NGC 4649 \citep{strader06}.  The same trend can be seen for the GCS
around the giant Sa galaxy NGC 4594 \citep{spitler06}, and \citet{mieske06}
find evidence for a similar MMR in their extensive sample of Virgo Cluster
Survey galaxies by combining their GC data into groups by host galaxy luminosity.
As H06 discuss, a clear trace of this trend had already been found for the Fornax cD
NGC 1399 by \citet{ostrov98} and \citet{dirsch03}, who noted that the
bright end of the GC population appeared broadly unimodal rather than
bimodal. As we now realize, this effect comes about
essentially because the blue and red sequences merge at the top end.

Intriguingly, the more metal-rich ``red'' sequence is not seen to exhibit
a MMR of its own in any galaxy so far.  That is, for the more metal-rich GCs,
metallicity is independent of mass.  The basic interpretation proposed
by H06 is that the blue-sequence clusters formed primarily in dwarf-sized
pregalactic clouds while the first stage of hierarchical merging was
beginning.  In these sites, more self-enrichment could occur within
the more massive dwarfs where more of the enriched gas from supernova
ejecta could be held back.  These same higher-mass dwarfs would also,
on average,
produce the most massive GCs \citep{hp94}.  By contrast, the more
metal-rich GCs on the red sequence formed later, perhaps in major mergers,
and at a stage when all the material could be held in the much deeper
potential well of the final big galaxy.  In that stage, enrichment would
have been more independent of GC mass.  \citet{mieske06} discuss a 
similar view.  \citet{rothberg07} use semi-analytic galaxy formation
models to test a different approach, namely the contribution of stripped
dwarf nuclei to the MMR.
They find that a MMR along the blue sequence
appears naturally with a standard model prescription for star formation
and enrichment rates in semi-analytic models.  An integral part of their interpretation is that
many of the massive blue GCs are actually the dense nuclei
of stripped dwarfs (dE,N or UCD objects).

Still more intriguingly, at least one clear anomaly
is already known to exist:  although the MMR along the blue sequence 
has now been found unequivocally in many galaxies, it does not seem
to occur in the well studied Virgo giant NGC 4472, where both the
red and blue GC sequences run vertically.  This result has been
confirmed in several studies and in a variety of photometric
bandpasses
\citep{gei96,lee98,puzia99,rhode01,strader06,mieske06}.  \citep{rothberg07}
find that increasing the efficiency of supernova feedback for
heating in dwarfs can produce this effect, leaving the blue GC
sequence without a MMR.  But it is not yet clear what underlying
physical or environmental conditions would cause this to occur.
In short, the existence of the MMR is beginning to provide rich
new insights into the early formation era of large galaxies.

NGC 3311, the central cD galaxy in the Hydra cluster A1060, has
long been an attractive target for GCS studies because it is
relatively nearby ($d = 53$ Mpc for a Hydra redshift $cz \simeq 3900$ km s$^{-1}$
and $H_0 = 70$ km s$^{-1}$ Mpc$^{-1}$) and has an enormously 
populous GCS comparable with the largest known, such as M87 in Virgo
or NGC 4874 in Coma.  Its ranking among the high-$S_N$ systems
was established with the early photometric survey by \citet{hsm83}.
Further ground-based imaging in the Washington system \citep{secker95,mcl95}
indicated that its GCS was moderately metal-rich with a hint
of bimodality, and that its spatial distribution was similar to the
shape of the cD envelope light.  Hubble Space Telescope imaging with
the WFPC2 camera \citep{brodie00} in the frequently used $(V-I)$ color
index was employed to argue that the MDF for its clusters was more like the ``standard''
pattern in giant ellipticals -- that is, a probable bimodal structure
with metal-poor and metal-rich sequences at their usual mean metallicities.
The most recently published study \citep{hempel05} combined the {\sl WFPC2}
$(V,I)$ data with {\sl NICMOS} imaging in F160W (approximately the $H-$band)
to gain a first rough estimate of the metallicity and age distributions
through models in the $(V-I, V-H)$ two-color diagram.

None of these earlier studies have given a satisfactory picture of the
system.  The HST-based studies \citep{brodie00,hempel05} are hampered by
the small fields of view of WFPC2 and NICMOS, yielding very
incomplete spatial coverage of this big cD galaxy,
and thus relatively small GC samples to work with.
By contrast, the initial study of \citet{hsm83} in the pre-CCD photographic
era covered a wide field but
penetrated to only quite shallow photometric limits by today's standards.
Relatively shallow limits and photometric calibration problems also affected
the first CCD-based imaging \citep{secker95}.
To study this unusually attractive GCS further, we need to draw from
a new photometric study that combines depth with wide field.

\section{Observations and Data Reduction}
\label{observations}

As part of a new program (Gemini GS-2006A-Q-24) intended to obtain deep multicolor
imaging of globular cluster systems around cD galaxies, 
we obtained ($g',i'$) images of NGC~3311 and the central Hydra 
cluster field using the GMOS imager
on Gemini South, which has a $5.5\arcmin \times 5.5\arcmin$ field of
view (FOV) and a scale of
0.146$\arcsec/pix$ after $2\times2$ binning.  The raw observations
were taken on the nights of 2006 February 8 and
March 23, under dark skies and photometric conditions, with an
average seeing of
$0.5\arcsec$.  In each waveband, 13 individual exposures were
taken, which we then reregistered and median-combined to create the final,
deep images in ($g', i'$).  The total integration time in each filter was
3900s.

The preprocessing of the raw exposures was
completed with the GEMINI package in
IRAF\footnote{IRAF is distributed by the National Optical Astronomy
Observatories, which are operated by the Association of Universities
for Research in Astronomy, Inc., under cooperative agreement with the
National Science Foundation.}.  In the $i'$ frames, a small amount
of night-sky fringing was present, which we 
successfully removed with a calibration fringe frame from the Gemini
archives.  After subtracting the fringe frame and then median combining 
all 13 $i'$-band frames, the resulting fringe signal contributes only 0.3 
percent to the background fluctuations.

%
%
\begin{figure} 
\figurenum{1} 
\plotone{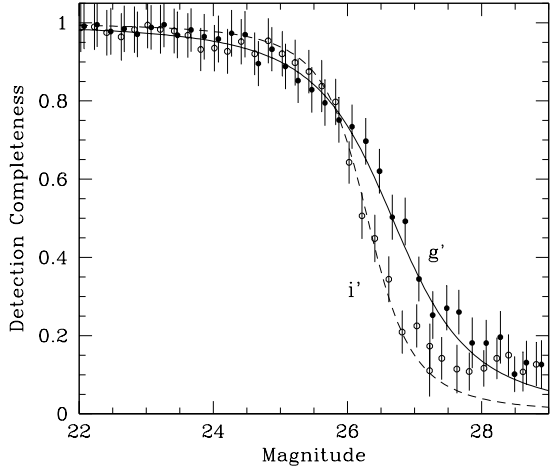} 
\caption{Photometric completeness of detection in $g'$ and $i'$.
Here the completeness fraction $f$, as determined from the artificial-star
tests described in the text, is plotted as a function of magnitude,
where $f(g')$ is plotted in the solid dots as a function of $g'$, and
$f(i')$ is plotted in the open symbols as a function of $i'$.
The interpolation curves drawn through each set of measurements are
the two-parameter functions listed in the text.
}
\label{completeness}
\end{figure}

For the photometric calibration, we used exposures of 
Landolt standard stars \citep{landolt92} taken adjacent to our
program-field images during photometric conditions.  We transformed the Landolt
standard-star indices in $UBVRI$ to ($g',i'$) using the
conversion equations of \citet{fukugita96}\footnote{A complete set of the
\citet{landolt92} standard stars transformed from the Johnson-Cousins
system into the Sloan filter set has been compiled by the authors and
is available online at
http://www.elizabethwehner.com/astro/sloan.html}. Since our standard
star exposures covered only a small range in airmass, we used 
the NGC 3311 images taken within just the same airmass range, and
taken during the same night within one hour of the standard-star exposures. 
This calibration was suitable only for establishing a mean zero-point 
value for our NGC 3311 frames, and not for a more complete evaluation
of the coefficients in airmass or color.  We therefore regard the
calibration as only a preliminary one, to be refined when a larger
network of standards can be obtained.

\begin{deluxetable}{ccccccccc}
\tablecaption{Basic Parameters for NGC 3311 \label{basics}}
\tablewidth{0pt}
\tablehead{
\colhead{Parameter} &  \colhead{Value} \\
}

\startdata
$Type$ & E2/cD \\
$\alpha$ (J2000) & $10^h 36^m 42\fs8$ \\
$\delta$ (J2000) & $-27\fdg 31' 42''$ \\
$v_r$ (helio) & 3593 km s$^{-1}$ \\
$A_V$ & 0.26 \\
$(m-M)_0$ & $33.62$ \\
$E(g'-i')$ & 0.158 \\
$V_T^0$ & 11.04 \\
$M_V^T$ & $-22.8$ \\

\enddata

\end{deluxetable}

The final measurement of our combined, deep ($g',i'$) images was
carried out with the latest implementation of the standalone
{\sl daophot} code, {\sl daophot4} \citep{harris07}.  We used
an extensive series of artificial-star tests with the {\sl addstar}
component of {\sl daophot} to evaluate the
internal photometric detection completeness as a function of magnitude:
stars were added to each the $g'$ and $i'$ images in groups of
500 over a wide range of magnitudes, 
the images were then remeasured in the same way
as the original frames,  and the fraction $f$ of stars recovered in each magnitude
bin was then determined.
The results are shown in Figure \ref{completeness}.
The limits of our data, defined as the magnitudes at
which $f$ drops to 0.5, are $g'(lim) = 26.67$ and $i'(lim) = 26.30$.  The trend of
$f$ with magnitude is well matched by a Pritchet interpolation function
\citep{fl95}, 
\begin{equation}
f = \frac{1}{2} \left[{1 -
         \frac{{\alpha}(m-m_{lim})}{\sqrt{1+{\alpha}^2(m-m_{lim})^2}}}\right].
\end{equation}
which has two free parameters: the limiting magnitude $m_{lim}$,
and a parameter $\alpha$ giving the steepness of the dropoff.  For
these images we find $\alpha_{g'}=0.8$ and $\alpha_{i'}=1.4$.
As will be seen below, the limiting magnitudes 
$g'(lim) = 26.7$ and $i'(lim) = 26.3$ are deep enough to
be comparable with the GC luminosity function turnover point of NGC 3311.
Within the bright bulge of NGC 3311, the photometry depth diminishes
due to the increased local sky noise at small radii.  However, for 
radii larger than $0\farcm6$ from NGC 3311, which include the vast majority of our 
data,  we find no change in the completeness limits $g', i'(lim)$.

The random uncertainties of the photometry, as determined from
{\sl daophot/allstar} and from the same artificial-star tests,
are well approximated by the interpolation curves
\begin{eqnarray}
\sigma(g')  =  0.01  +  0.050 ~{\rm exp} (g'-26.0), \\
\sigma(i')  =  0.01  +  0.050 ~{\rm exp} (i'-25.0).
\end{eqnarray} 
They show little variation from place to place within the GMOS
field, indicating that crowding is unimportant, and that
background light from NGC 3311 and NGC 3309
has little influence except near their very centers.
On the basis of these error curves, we would expect the GC
sequences to start being noticeably `broadened' in color,
purely because of random uncertainties in the photometry,
for levels $i' \gtrsim 24$.  Lastly, the artificial-star tests show
that any {\sl systematic} error (net bias) in the photometry is
less than 0.02 mag in each filter for $i' < 25, g'< 26$.

Throughout the following discussion we adopt a distance modulus $(m-M)_0 = 33.62$
for NGC 3311 and foreground
reddening $E(g'-i') = 0.158$, as we did in \citet{wehner07}.

\section{The Color-Magnitude Distribution}
\label{CMD}

After carrying out the photometry in $g'$ and $i'$ with
{\sl daophot} we used the additional code SExtractor
\citep{bertin} to help reject obvious nonstellar
objects from the detection lists (at $0.5''$ seeing and at the
distance of the Hydra cluster, globular clusters appear 
indistinguishable from the starlike PSF.  Under HST-type resolution
of $0.1''$, the profiles of the 
very biggest GCs can, however, be resolved; see Wehner \& Harris 2007).
Detections from the two wavebands were then
correlated, and only those objects found in both were kept
for the final analysis.   

The final color-magnitude measurements for 8108 starlike objects
across the entire GMOS field
are shown in Figure \ref{cmd}.  The globular cluster population is
readily visible as the dominant concentration of points in the color range
$0.6 \lesssim (g'-i') \lesssim 1.4$.  Fainter than $i' \sim 25$
the field contamination and photometric scatter increase markedly,
but brighter than this, the GC population is well defined.  Evidence
of the standard ``blue'' and ``red'' sequences can already be seen
and will be discussed more quantitatively below through objective
bimodal fits.  On the red-sequence side (approximately in the range
$1.0 < (g'-i') < 1.3$) a distinctive bright extension to high
luminosity can be seen that, intriguingly, does not exist for the
blue GC sequence.  This bright-end  feature extends up into the
GC mass range $M \gtrsim 10^7 M_{\odot}$ that overlaps with the
regime of the ``Ultra-Compact Dwarfs'' (UCDs).  We have discussed
these objects in our previous paper \citep{wehner07}.

To date, not much other photometry of globular cluster systems
in the $(g', i')$ indices has
been published that we can make comparisons with.  
\citet{forbes04} present $(g'-i')$ data 
taken with the GMOS camera on Gemini North for the Virgo giant
NGC 4649.  The bimodal fit they obtain to its GC color distribution
has mean values $(g'-i')_0$ for the blue
and red sequences which are $\sim 0.1$ mag {\sl redder} than our
calibrated and dereddened colors for NGC 3311.  We would expect
{\sl a priori}, however, that the colors of the two sequences
would be quite similar from one galaxy to the next
\citep[e.g.][]{har06,strader04} with only a modest trend
due to the parent galaxy luminosity.  For the present,
we do not regard this difference as significant, because
(as emphasized above) we regard our zeropoint 
calibration as a preliminary
one, which still needs a separate series of observations especially
to establish the photometric color terms.
Notably, Forbes et al. also regard their own zeropoint calibration
as preliminary.
What is perhaps more important at this stage, and equally worth
noting, is that the mean color {\sl difference} 
$\Delta (g'-i') \simeq 0.30 \pm 0.08$ between the
red and blue sequences that we find for NGC 3311 (see the 
discussion below) is identical with
the separation btween those two sequences measured 
for NGC 4649 by \citet{forbes04}.

\section{The Global Features of the Globular Cluster System}
\label{NGC3309}

In sections 4.1 -- 4.2 we use our GMOS data to discuss the
spatial distribution of the GCs around both NGC 3311 and 3309,
including the radial and azimuthal components.  A continuing
theme in this analysis is the relative contribution of the two
galaxies to the GC population we see on our images.  
In section 4.3, we discuss the color (metallicity) distribution
of the GCs by radius around NGC 3311, and in sections 4.4 -- 4.5
we discuss the GC luminosity function and specific frequencies.

\subsection{Measuring the Radial Profiles:  The NGC 3309 Problem}

The core of the Hydra cluster contains several other large galaxies
in addition to the central supergiant that was
our primary target.  Of these others, the closest
in projection and by far the
most luminous is the giant elliptical NGC 3309, just $100''$ away from 
the center of NGC 3311.  Its integrated magnitude ($V_T = 11.94$; see below)
places it 0.9 magnitude fainter than NGC 3311
($V_T = 11.04$), though the true total luminosity of NGC 3311 is estimated
to be about 1.3 mag brighter than that of NGC 3309 
because of its extended cD envelope.  Nevertheless, we might
expect that NGC 3309 would contribute significant numbers of GCs of
its own to the population of objects in the field. 
By using starcounts around each of the two galaxies on the sides opposite
to the other galaxy, \citet{hsm83} and \citet{mcl95} both argued that NGC 3309
was likely to be contributing no more than a few percent of the
GCs in the Hydra core field, with NGC 3311 acting as the
dominant contributor.  Even though NGC 3311 is clearly a high-$S_N$
elliptical like many cDs, such claims might mean that NGC 3309 would
have unusually low $S_N$ for a giant elliptical in a rich cluster.
A better and more quantitative estimate of the relative populations,
as well as the spatial structures of the two GCSs,
is possible from our deeper data.

To quantify our analysis, let us assume that the projected number density of
GCs on the sky from each of the two galaxies
can be modelled by a circularly symmetric function $\sigma(r)$
giving the number of GCs per unit projected area
(we justify the assumption of symmetry below).  Then, the
total number density of counted objects at a point $(x,y)$
anywhere in the field is given by
\begin{equation}
\sigma(x,y) \, = \, \sigma_b \, + \, \sigma_1(r_1) \, + \, \sigma_2(r_2)
\label{fiteqn}
\end{equation}
where $r_1$ is the distance of the given point $(x,y)$ from the
center of NGC 3311, $r_2$ is its distance from the center of NGC 3309,
and $\sigma_1, \sigma_2$ are the two radial functions specifying the
GCS around each galaxy.  Finally, $\sigma_b$ is the uniform background density
of whatever contaminating objects fall within our selected sample.
Our goal is to deduce, from a fit to the observed GC distribution in
the field, the best-estimate parameters for $\sigma_1$ and $\sigma_2$
{\sl simultaneously}.

To perform the fit, we adopt two possible models for the distribution:

\noindent (a) A standard Hubble profile,
\begin{equation}
\sigma(r) \, = \, b \bigl(1 + {r \over c} \bigr)^{-a}
\label{hubble}
\end{equation}
which is a power law with an added core profile of core radius $c$.

\noindent (b) A generalized Sersic profile,
\begin{equation}
\sigma(r) \, = \, \sigma_e {\rm exp}\Bigl[-b_n \bigl(({r \over r_e})^{1/n} - 1 \bigr) \Bigr]
\end{equation}
where $b_n \simeq 1.99 n - 0.3271$ \citep{ferr06} to guarantee that
$r_e$ will be the half-light radius, and for typical
E galaxies, the exponent $n$ is found observationally to 
fall in the range $\simeq 1 - 4$.
The case $n=4$ recovers the standard de Vaucouleurs law.

\begin{figure} 
\figurenum{2} 
\plotone{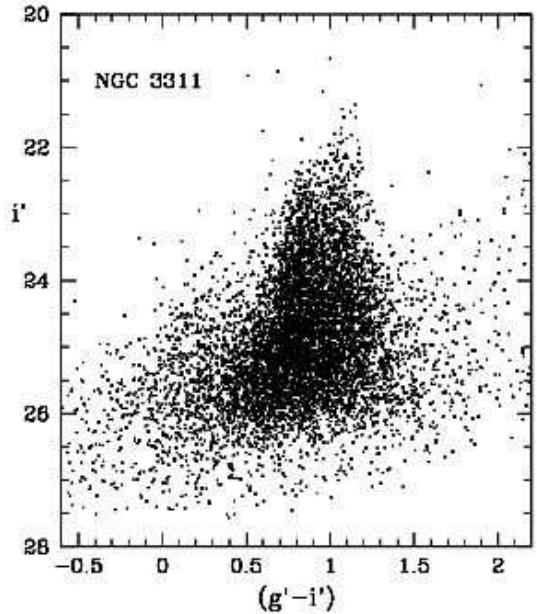} 
\caption{The color-magnitude diagram for all measured
objects in the GMOS field centered on NGC 3311.  
}
\label{cmd}
\end{figure}

\begin{figure} 
\figurenum{3} 
\plotone{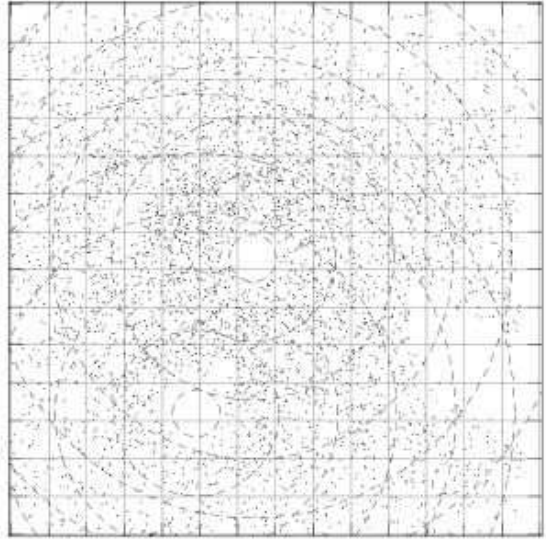} 
\caption{Grid layout showing the determination of the
spatial structure of the GCS.  Here North is at left
and East at top, following the instrumental orientation
of the GMOS camera during the observations.  The center
of NGC 3311 is near the center of the field, while the
center of NGC 3309 is at lower left; the guide probe
shadow is at lower right.  The dots are the measured
objects in the restricted range $(i' < 25.0, 0.6 < (g'-i') < 1.4$
as described in the text.  Concentric circles are drawn
around each of the two major galaxies, while the grid
lines show the $14 \times 14$ matrix of boxes used to
solve simultaneously for the radial distributions of
the GCSs around both galaxies.
}
\label{grid}
\end{figure}

To carry out the actual numerical fit, we divide up the observed area
into $k \times k$ boxes, each of area $A$; we use the model parameters to predict how many
objects $N_{ij} = A \cdot \sigma(x_i,y_i)$
should be found in each box at position $(x_i, y_j), i,j = 1, ... k$;
and then calculate the residuals (observed $-$ predicted) in each box.
A standard goodness of fit based on Poisson statistics can then be formed,
\begin{equation}
\chi^2 \, = \, \sum {{\bigl(N_{ij}(obs) - N_{ij}(pred) \bigr)^2} \over {N_{ij}(pred)}}
\end{equation}
The outline of the model is shown in Figure \ref{grid} for the $14 \times 14$
grid that we used for the final reductions.  An advantage of this procedure is
that it uses the information everywhere in the frame, not just the counts within
some arbitrary radius centered on NGC 3311.  In practice, we ignored any
boxes located at the centers of the two galaxies and at one very
bright foreground star, as well as the boxes under the shadow of the 
camera guide probe (lower right part of Fig.~\ref{grid}) and at the vignetted
corners of the field.  To reduce field contamination from the outset
(see Fig.~\ref{cmd}), we used the sample of 4393 objects within the restricted range 
$i' < 25.4, 0.6 < (g'-i') < 1.4$, expecting that this sample would be
dominated by GCs.

When we fit Eq.~\ref{fiteqn} to the data, in principle we have 
a total of six free parameters.
For the Hubble profile model, these six are 
$\sigma_b, a_1, c_1, a_2, c_2$, and the relative amplitude $b_1/b_2$.  
For the Sersic model, they are $\sigma_b, r_{e1}, r_{e2}, n_1, n_2$ and 
$\sigma_{e1}/\sigma_{e2}$.
Once a particular ratio $b_1/b_2$  
is chosen, the actual values of both $b_1$ and $b_2$ individually
are automatically set by the
condition that the {\sl total} predicted population of objects in the field
must precisely equal the observed total.
The same applies to $\sigma_{e1}$ and $\sigma_{e2}$ for the Sersic model.

In practice, the background density $\sigma_b$ is extremely hard to
determine by the fitting procedure alone, because the NGC 3311 GCS is
so extensive that it dominates the starcounts even near the edges of
our GMOS field.  An external constraint on $\sigma_b$ is therefore
desirable.  Lacking any remote ``control field'' to rely on,
we use the distribution of stars in the CMD itself 
(Fig.~\ref{cmd}) to estimate the background.
The background contamination will be due to a combination of
foreground Milky Way stars and faint, small remote galaxies.
We can estimate the number and color distribution of possible field
stars using the galactic stellar population synthesis models\footnote{http://bison.obs-besancon.fr/modele/} of 
\citet{robin03}.  Using a $5.5\arcmin \times 5.5\arcmin$ field centered on
NGC 3311, the models find only 13 stars, suggesting contamination from
galactic foreground stars is small at these latitudes.  Using a larger
field of view, such as a square degree, results in over 2 thousand stars 
with a color distribution leaning blueward of the blue peak of the GCS 
distribution.  Our main source of contamination, as indicated
by the initial FOV search, is most likely distant background galaxies.
Without a control field, we cannot specifically quantify these background galaxies.  
If we assume that their distribution of colors is roughly uniform
over the interval $0.0 \lesssim (g'-i') \lesssim 2.0$, we can
use the numbers of objects with colors in the ranges
$0.0 < (g'-i') < 0.6$ and $1.4 < (g'-i') < 2.0$ to predict how
many background objects are in our selected midrange
of $0.6 < (g'-i') < 1.4$.  For $i' < 25.4$ we count
860 objects total that are bluer or redder than the GC color range.
Normalizing to the same color interval, we then estimate that 570 of
the 4393 objects within $0.6 < (g'-i') < 1.4$ (or about 13\%)
are background contamination, equivalent to $\sigma_b = 18$ arcmin$^{-2}$.
In the numerical fitting procedure we adopted this background
throughout and treated only the 5 remaining parameters in each model as free.

For the Hubble profile model, we obtain a best-fit set of parameters
$a_1 = -2.04 \pm 0.07$ and $c_1 = 94'' \pm 8''$ for NGC 3311; and
$a_2 = -2.8 (+0.7, -1.3)$ and $c_2 = 55'' (+30'',-24'')$ for NGC 3309.
The ratio $(b_2/b_1)$ is $ 0.093 \pm 0.057$. 
For this set of parameters the minimum
reduced $\chi_{\nu}$ value was 1.44.  We experimented with
different grid sizes and found that the results were not sensitive
within broad limits to the fineness $k^2$ of the chosen grid.
Given the simple nature of the model, and other factors not accounted
for such as the necessarily patchy sampling across the grid, the
internally uncertain adopted level of the background, and
possible clumpiness (clustering) in the background population,
the fit is successful at accomplishing our main goal to evaluate
the relative contributions of the two giant galaxies.

A very slightly better overall fit (minimum $\chi_{\nu} = 1.41$) 
is obtained with the generalized
Sersic model.  For the NGC 3311 GCS, we find
$n_1=1.24 \pm 0.06$ and $r_{e1} = 177'' \pm 9''$, while for NGC 3309
$n_2 = 1.1 (+1.2, -1)$ and $r_{e2} = 93'' (+78'', -45'')$.  The ratio
of the two amplitudes is $(\sigma_{e2}/\sigma_{e1}) = 0.080 \pm 0.046$.

For comparison, the scale sizes of the underlying halo {\sl light}
of each galaxy are significantly smaller:  \citet{vasterberg91} find
$r_e \simeq 21\farcs5 \simeq 5.6$ kpc for NGC 3309 and 
$94\farcs2 \simeq$ 25 kpc for NGC 3311
from an integrated-light study in the Gunn $r$ filter.  These effective
radii are not strictly comparable to our solutions for the GCSs,
because they assume a classic de Vaucouleurs profile ($n=4$ in the 
Sersic model); but they confirm that the GCS is spatially more
extended, as has usually been found for giant galaxies
\citep{mcl99,harris01}.

One very evident result of this numerical exercise is that
NGC 3309 contributes only a minor part of the GC population in our
field (2 to 3 percent; see section 4.5 below). 
As a consequence, its GCS shape parameters are quite weakly
constrained.  Assuming that the
detected objects in the field belong {\sl entirely} to NGC 3311
and ignoring NGC 3309 completely (that is, setting $\sigma_2 \equiv 0$ in
the numerical fitting) yields a minimum $\chi^2$ negligibly
worse than the two-galaxy fit.   We quantify the
relative importance of the two galaxies a bit further below.

\begin{figure} 
\figurenum{4} 
\plotone{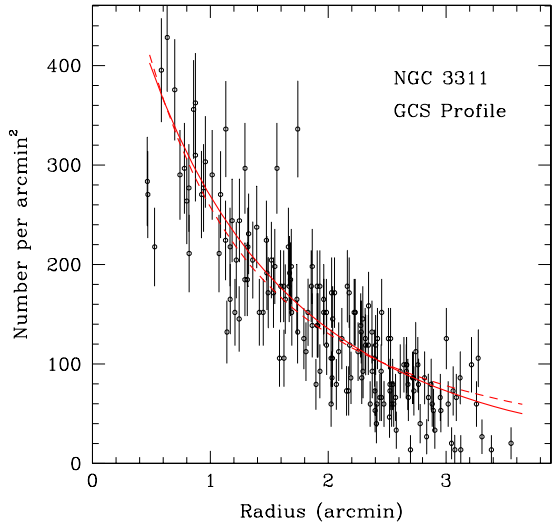} 
\caption{Radial distribution of the GCs in the field
relative to the center of NGC 3311.  The number of objects
within each of the grid boxes shown in the previous figure
is plotted against the radial distance of the box center
from NGC 3311.  The best-fitting Hubble profile (see text)
is shown as the dashed line, while the best-fitting 
generalized Sersic model is the solid line.
}
\label{radial}
\end{figure}

A rough comparison of the two model fits is shown in
Figure \ref{radial}, where we simply plot the number density of objects
in each grid box as a function of distance from the center of
NGC 3311.  The best-fit Hubble and Sersic models {\sl for NGC 3311 alone}
are shown overplotted, verifying the first-order description of
the entire population in the field as belonging to the central cD.
The same grid plot, but this time centered on NGC 3309, gives a radically
different impression, similar to picking just a random point somewhere
away from NGC 3311.  That is, no significant concentration towards
NGC 3309 is evident.

A significant improvement in this estimate of the two GCS spatial
structures will be possible if a better external measurement of
the background can be obtained.  The NGC 3311 system is so extensive,
however, that photometry at projected radii $r \gtrsim 10'$ (150 kpc)
will be necessary for a control-field location.

\subsection{Azimuthal Distributions}

\begin{figure} 
\figurenum{5} 
\plotone{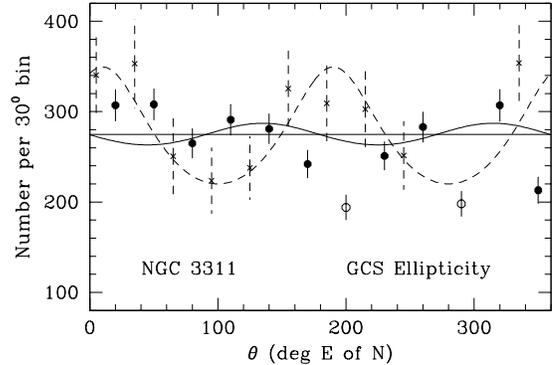} 
\caption{Azimuthal distribution of the GCS around NGC 3311.
The numbers of objects within 30-degree sectors, and within
radii $0\farcm26 < r < 2\farcm50$ of NGC 3311, are shown
as the data points with error bars.  The total sample includes
3544 objects divided into 30-degree sectors;
the two open circles denote the sectors containing the center
of NGC 3309 and the shadow of the GMOS guide probe.
These are not used in the numerical fit.
The sinusoid-like solid curve (see the text for the actual
model curve) is the best-fitting ellipsoidal model, with
$\epsilon = 0.07$ and major axis $\theta_p = 90^o$.
For comparison, the solution obtained by
McLaughlin et al.~(1995) from a smaller photometric sample and
a brighter limit is shown by the crosses with error bars and the dashed curve.
}
\label{azimuth}
\end{figure}

\citet{mcl95} also attempted to measure the ellipticity  of
the NGC 3311 GCS and to compare it with the shape of its cD envelope.
We can, in principle, make a better test with our much larger measured
population.  In this case, for our test sample we take all 3544
measured objects within radii $0\farcm26 < r < 2\farcm50$ of NGC 3311
(the outer circle is the largest complete one fitting within 
the GMOS frame boundary) and within the photometric limits
$i' < 25.4, 0.6 < (g'-i') < 1.4$.  This sample is subdivided into
$30^o$ sectors and plotted against azimuthal angle $\theta$ of the
sector, as shown in Figure \ref{azimuth}.
The two sectors containing the center of NGC 3309 and
the guide probe shadow are shown as open symbols in Fig.~\ref{azimuth}
and are ignored.

It is evident that there is no
strong trace of ellipticity in the residuals about
the mean level of $\simeq 275$ objects per sector after background subtraction.
Although there are sector-to-sector differences larger than
the internal statistical uncertainties, they cannot easily be made to
match an approximate double-cosine curve that would
be the mark of a modestly elliptical distribution
\citep[][]{mcl94,mcl95}.
The integrated halo light of NGC 3311 has a small but definite 
ellipticity of $\epsilon = 0.2$ and major axis at $\simeq 30^o/210^o$
\citep{disney77,vasterberg91}.  If the formation of the GCS is 
closely associated with the galaxy halo, we should expect no large
differences in their intrinsic shapes and orientations.  Contrarily,
a dramatic difference in the GCS shape or orientation relative to
the halo light would be a signal of a more fundamental difference
in the GCS formation era or its internal dynamical evolution.

We have carried out a solution for $\epsilon$ and $\theta_p$ with
our GCS data, using the formalism of \citet{mcl94,mcl95}.
They show that
for an intrinsically elliptical distribution of points sampled in
{\sl circular} annuli, the observed number density is
\begin{equation}
\sigma(r,\theta) \, = \, \sigma_b \, + \, \sigma_0 r^{-\alpha}\bigl[{\rm cos}^2 (\theta-\theta_p)
  + (1 - \epsilon)^{-2} {\rm sin}^2 (\theta - \theta_p) \bigr]^{-\alpha / 2}
\end{equation}
where the radial falloff is modelled as a simple power law of exponent $\alpha$.
Here, $\theta_p \pm 180^o$ denotes the position angle 
of the major axis at which $\sigma(r,\theta)$ reaches its maximum value.
For NGC 3311, we have $\alpha \simeq 1.2$ (for the Hubble profile fit
described above, we found $\alpha = 2.0$, but if we remove the core-radius
parameter $c$, we require a flatter overall power law to match the whole profile
adequately).  We then vary the assumed $\epsilon$ and $\theta_p$ till
the residuals about this model are minimized.  

The best-fit solution we
find is for $\epsilon = 0.07 (+0.11, -0.07)$ and $ \theta_p = 90^o \pm 50^o$,
shown as the solid curve in Fig.~\ref{azimuth}.  The corresponding major axis
line is thus $90^o/270^o \pm 50^o$.
(In Fig.~\ref{grid}, the major axis would run vertically.)
The best-fit solution is weakly determined with 
a shallow minimum at $\chi_{\nu} = 1.95$, and indeed
solutions with $\epsilon \simeq 0$ are
scarcely worse than this one.  Our deduced ellipticity is
lower than for the integrated light, and 
our solution for $\theta_p$ falls $\simeq 60^o$ eastward of the
integrated halo light, though its uncertainty is large enough that
there is no formal disagreement.  In short,
we appear to be confronted with no difficulties by assuming
first-order circular symmetry.

For comparison, we show one of the solutions by
\citet{mcl95} as the dashed line in Fig.~\ref{azimuth}; 
this clearly does not match our solution very well.
\citet{mcl95} performed solutions for two different magnitude
ranges ($T_1 \leq 23$, $T_1 \leq 24$)
and for the more slightly more restricted radial range 
$0\farcm89< r < 2\farcm22$.    For $T_1 \leq 23$, they found
$\epsilon = 0.32 \pm 0.05$, $\theta_p = 145^o$, while for
$T_1 \leq 24$ they found 
$\epsilon \simeq 0.3$, $\theta_p = 190^o$ (this latter one is shown in our
Fig.~\ref{azimuth}).  They comment that ``the uncertainty in $\theta_p$
is at least $\pm 40^o$''.

Both of the McLaughlin et al. solutions differ from ours;
we find a distinctly smaller ellipticity and a position angle
between their two solutions (although, again, 
it is not clear that any genuine formal disagreement exists).
To attempt to trace down causes for the evident differences,
we rebinned our data to magnitude ranges corresponding more closely
to theirs, and also restricted to exactly the same radial range as
they used.  For these numerical trials we also included objects
at all colors, as they did.  For our objects brighter than $i' = 23$,
we find $\epsilon = 0.34 \pm 0.15$ and $\theta_p \simeq 110^o/290^o$.
For a slightly deeper cut down to $i'=24$, we find
$\epsilon = 0.10 \pm 0.07$ and $\theta_p \simeq 75^o/255^o$.
Interestingly, the best-fit position angles of our two subsamples
bracket our best solution from the large, `best' dataset ($i' < 25.4$
and color-selected), but restricting the sample much more severely
to the brightest objects tends to lead to a larger $\epsilon$, as
found by McLaughlin et al.
Without being able to offer definitive reasons
for the disagreement between these two studies, we suggest that the
earlier GCS result was likely to 
have been compromised by the combination of small-number statistics,
and especially by much larger relative field contamination
(since the dominant source of field contamination is faint, small
background galaxies which have clumpy space distributions, they
are quite able to bias our solutions for the spatial distribution
if they contribute a high fraction of the total sample).
In the present data, we have been able to reject non-GC objects
more completely by color and morphology than in the older study.

The most intriguing feature of the azimuthal distribution is the
residual difference between the GC system ($\epsilon \sim 0.1$,
$\theta_p \sim 90^o$) and the halo {\sl light} ($\epsilon \sim 0.2$,
$\theta_p \sim 30^o$).  Unfortunately, we are not yet in a position
to place heavy significance on this, simply because the solution for
the best-fit position angle is extremely uncertain if the 
ellipticity is very small.  If the difference is verified with
deeper and cleaner samples, it will, perhaps, be a tantalizing hint
that the GC accretion or destruction histories differed in
second-order ways from the field stars.

\subsection{Radial Color Distribution}

With the knowledge that the vast majority of GCs in this field belong
to NGC 3311, we can also investigate their color distribution as a function
of galactocentric distance.  In many GCSs within large ellipticals, a color gradient can be found
in the sense that the mean color of the GCs (averaging together both the red
and blue ones) conventionally 
decreases with increasing $R_{gc}$.  However, beginning especially with the
key work of \citet{gei96} on NGC 4472, it became clear that (in Geisler et al.'s
concise wording)
``most of this gradient appears to be due to the varying radial concentrations
of the two populations''.  That is, the metal-poor, blue GC population is less
centrally concentrated than the metal-richer population
and thus contributes a higher proportion of clusters at larger radius,
generating an overall declining metallicity gradient in
the GC population as a whole.  Notably,
each of the two metallicity components shows little or no color (metallicity)
gradient of its own.  Very much the same conclusion had already
become clear long ago (albeit with a much smaller sample of GCs than can be found
in giant ellipticals) for the Milky Way GCS \citep{zinn85}.

\begin{figure}
\figurenum{6}
\plotone{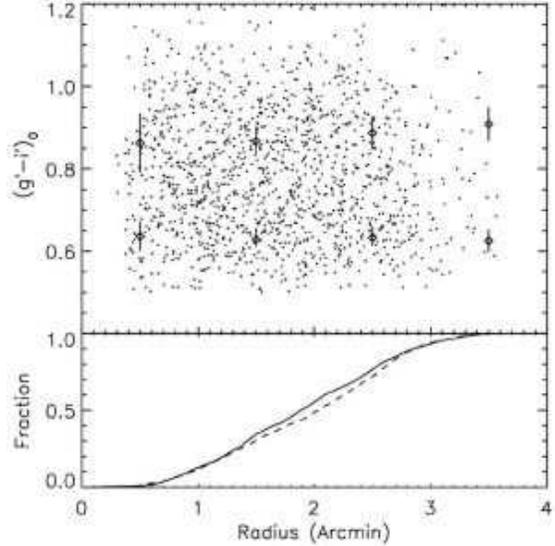}
\caption{{\sl Upper panel:} Color index $(g'-i')_0$ of globular clusters in our GMOS
field, plotted as a function of radial distance from NGC 3311.
All objects brighter than $i = 24.0$ are included here.  The overlaid diamonds indicate the
RMIX fits for the blue and red peaks of the distribution.    
{\sl Lower panel:} Cumulative radial distributions of the metal-poor clusters
(dashed line) and metal-rich clusters (solid line).  The lower-metallicity
clusters are slightly more extended spatially than the metal-richer ones,
but not significantly so.
}
\label{radcolor}
\end{figure}

In Figure \ref{radcolor} we show the color indices for the objects in the
NGC 3311 field (in order to minimize field contamination, we restrict the
sample to the 1360 objects brighter than $i'=24.0$).  A broad range in color
is present at all radii, but no strong
radial gradient in color is evident for either the red or blue subpopulations,
a similar result to other galaxies.  In order to determine whether there was a 
trend in color versus radius, we binned the clusters
in radius and performed a fit using RMIX (a fitting code discussed in detail in section 6).  The locations of the red and blue peaks are
shown as diamonds in Figure \ref{radcolor}.  A comparison between the cumulative distributions
of the red and blue clusters (shown in the lower panel of the Figure) indicates that
any difference has weak significance, at least within the regions our data
are sampling.  Unfortunately, the field size we sample
reaches a maximum $R_{gc} \sim 50$ kpc, which is not large enough to test
either the true spatial extent of the system or what happens to the color
distribution near its outskirts (recall, for example, that the M87 GCS
extends detectably outward to at least $R_{gc} \simeq 100$ kpc; see \citet{tam06}).

\subsection{Luminosity Function}

Deep imaging from galaxies of all types now shows that the luminosity
distribution of their GCs follows a roughly universal shape that has
most often been approximated by a Gaussian distribution in number 
of GCs per unit magnitude 
\citep[e.g.][among many others]{harris01,kundu01a,kundu01b,larsen01,jordan07}.
For giant galaxies, the turnover or peak frequency of the GCLF
is consistently near $M_V^0 = -7.4 \pm 0.2$ \citep{harris01}
while the intrinsic dispersion is in the range $\sigma = 1.3 - 1.5$
\citep[see particularly][for an extensive analysis of the correlation
of the turnover and dispersion with host galaxy luminosity, although it
is worth noting that the
only cD-type galaxy included in their data is M87]{jordan07}.

Ideally, a determination of the GCLF for our target NGC 3311 would best
be done by taking the LF of an adjacent ``control field'', observed
under identical conditions, and subtracting it
from the LF of the objects in our $5\farcm5-$wide
central field.  Lacking this, a second-best alternative is to use the
outer parts of our GMOS field as a ``background''.  In fact, much of the
outer zone is still populated by GCs, and so 
by defining it as background we will
partially subtract some of the GC population as well as the true background.
However, as long as no large changes in the GCLF with radius exist, the
residual GCLF we obtain this way will be systematically correct
even if a bit oversubtracted.

An inevitable limitation of using the outer parts of the image field
to define the background level is that the background estimate $\sigma_b$
is much more uncertain numerically than if we had a large, remote control
field to work with.  
We divide our field into two regions:
an inner annulus of inner radius $r_0$, outer radius $r_1$, and area $A_{in}$; 
and an outer zone of inner radius $r_1$ and outer boundary set by the edge
of the frame, and area $A_{out}$.  We set $r_0$ to be the innermost radius at which the photometry
is still highly complete to the magnitude limit of interest ($i' \simeq 26$).
The only significant free parameter we have to define these two zones is the
mid-radius $r_1$.  The residual GC population within $A_{in}$ after background
subtraction will clearly be
\begin{equation}
N_r \, = \, N_{in} - (A_{in}/A_{out}) N_{out}
\end{equation}
where, by hypothesis, the background $\sigma_b$ is constant across the field
and subtracts cleanly out after the area normalization.
We would like to choose the midpoint $r_1$ in such a way as to minimize the relative
uncertainty in $\sigma(N_r)/N_r$.  If $r_1$ is too large, then $N_r$ will
be large, but the area
normalization factor $(A_{in}/A_{out})$ will blow up and the background
correction becomes extremely uncertain.  But if $r_1$ is too small, then so is $N_{in}$,
and again the relative uncertainty in the residual population becomes large.

To fix $r_1$, we adopted the radial profile parameters described above for
the GC system (the best-fit Hubble profile plus the background 
level $\sigma_b = 18$ arcmin$^{-2}$).  The inner radius $r_0$ is set at $0\farcm5$.
We then calculated the resulting relative uncertainty $\sigma(N_r)/N_r$ as a
function of $r_1$, where we let $r_1$ vary from just slightly larger than $r_0$
out to almost the borders of the field.  The optimum $r_1$ turns out to be roughly
halfway from the center of NGC 3311 to the edge of the field, given the geometry
of this system, but the relative uncertainty goes through a very shallow minimum
near this point and the precise choice is not critical.  For the following
discussion, we use $r_1 = 2\farcm5$.
The ``background'' region is defined to be everything 
beyond that radius, though not including the vignetted regions in
the outskirts.
In each zone we count all objects within the broad color range $0.2 < (g'-i') < 1.4$,
and also correct the raw numbers of objects in each bin
for photometric incompleteness $f = f_{i'} \cdot f_{g'}$
with the completeness function parameters derived above.  

\begin{figure} 
\figurenum{7} 
\plotone{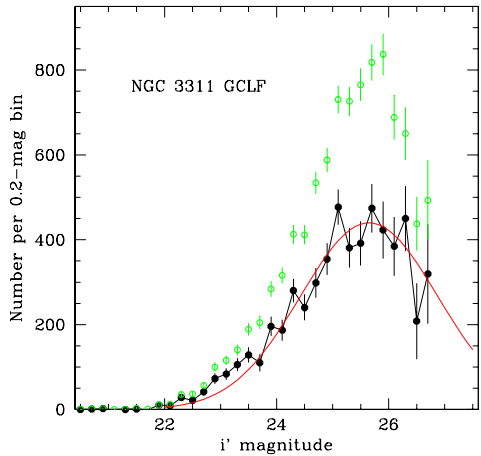} 
\caption{Luminosity function for the globular clusters in
NGC 3311. The {\sl open circles} with errorbars show the
numbers of objects within the radial range $0\farcm5 < r < 2\farcm5$.
The connected {\sl solid dots} with errorbars are the totals
after approximate background subtraction, where the GMOS image
region outside the circle $r > 2\farcm5$ is used to define
the background.  The red solid line is a Gaussian LF with
peak at $i' = 25.65$ (corresponding to $M_V \simeq -7.4$)
and with standard deviation $\sigma_{i'} = 1.5$.
}
\label{gclf}
\end{figure}

The results are in Figure \ref{gclf},  shown
in $\Delta i' = 0.2-$mag bins.  For $i' > 26.5$,
the net completeness $f$ falls below 0.5 and the LF becomes considerably
more uncertain (note the rapid growth of the internal error bars).
This level is also very near the {\sl predicted} magnitude of the
GCLF turnover point for a normal giant or supergiant elliptical galaxy.  
For $M_V^0 \simeq -7.4$ and $\langle V-I \rangle_0 = 1.0$, we expect
$M_I^0 \simeq -8.4$ or $I^0 = 25.22$  adding the distance modulus $(m-M)_0 = 33.62$.
To convert $I$ to $i'$, we use
the mean difference $\langle i'-I \rangle = 0.28 \pm 0.02$
determined directly within our field
from 12 of the brightest GCs \citep[see][]{wehner07}.
Finally adding the foreground extinction
$A_I \simeq A_{i'} = 0.15$, we have $i'{\rm (turnover)} \simeq 25.65$.
A Gaussian curve with that peak magnitude and with $\sigma = 1.5$ mag
is shown superimposed on the observed GCLF in Fig.~\ref{gclf}.
Encouragingly, the raw LF of all objects within the inner zone
(shown as the open circles) has the same peak to within $\sim 0.1$ 
mag even without background subtraction.

The Gaussian curve shown in Fig.~\ref{gclf} 
is not intended to represent a rigorous derivation
of the observed turnover point or dispersion; we use it
only to show that the standard values provide an
entirely plausible description of the data for NGC 3311. That is, the
GCLF in this cD galaxy gives no indication of being unusual.  
Our photometry does not
reach far enough past the turnover to permit any kind of useful test
of more sophisticated numerical models, such as the asymmetric
Schechter-like function explored in depth by \citet{jordan07} from the
Virgo cluster survey.  However, in a broader sense our data clearly show that 
imaging with ground-based 8m-class telescopes is capable of
penetrating deep into the globular cluster populations of 
galaxies in the 50-Mpc distance range and even beyond.

Recent interest has followed the possibility that the GCLFs may differ
between the redder and bluer metallicity groups of clusters.  If the
underlying {\sl mass} distributions of both types of clusters are similar,
then the GCLF turnover magnitude should differ between photometric
bandpasses by typically $\sim 0.2$ mag purely because of differential bolometric
correction \citep{ashman95}.  Larger differences between the red and blue
GCLFs (or ones in the opposite sense expected from bolometric corrections)
may be indicators of genuine differences in their mass distributions.
In an HST-based survey of several nearby ellipticals, \citet{larsen01} found
a mean difference $\Delta M_V = 0.4$ mag in the sense that the blue GCLF peak
was brighter, though large and uncertain differences show up from one system
to another.  They conclude that it is unclear whether any true underlying
mass differences are required to explain this mean difference.

\begin{figure} 
\figurenum{8} 
\plotone{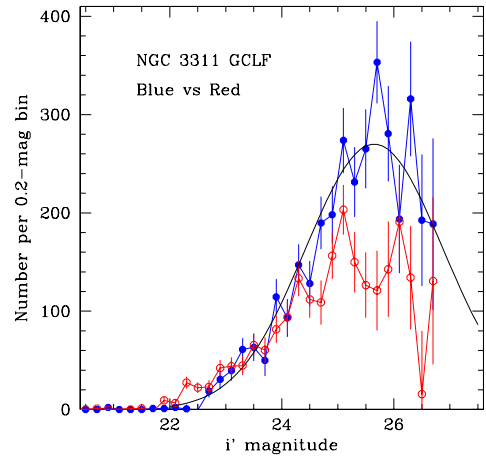} 
\caption{Luminosity function for the globular clusters in
 NGC 3311, plotted separately for the blue and red GC sequences.
The {\sl open circles} with errorbars show the redder clusters
lying in the color range $0.8 < (g'-i')_0 < 1.4$, while
the {\sl solid dots} with errorbars show the bluer clusters
in the color range $0.2 < (g'-i')_0 < 0.8$.  Both sets of points
have been background-subtracted as described in the previous figure.
The solid line is a Gaussian LF with the same mean and standard
deviation as in the previous figure, matching the entire GCLF
over all colors.
}
\label{gclf_color}
\end{figure}

NGC 3311 has large numbers of both red and blue clusters, and using the
same numerical technique as described above, we determined their GCLFs
separately.  ``Blue'', metal-poor clusters are defined to lie within
the color range $0.2 < (g'-i')_0 < 0.8$ and ``red'', metal-rich clusters
within $0.8 < (g'-i')_0 < 1.4$.  The results are shown in Figure \ref{gclf_color},
where we display only the residual LFs and not the raw numbers before
background subtraction.  The blue population falls close to the same
roughly-Gaussian curve describing the system as a whole.  For the red population,
the small excess of very luminous clusters $(i' < 23)$ is evident, but the
most striking difference is in the range $i' = 25 - 26$, where a noticeable
dip is seen just around the nominal turnover point.  We are, however, reluctant
to conclude that this effect is real given the uncertainties in the background
correction and the determination of residuals.  A standard Kolmogorov-Smirnov
two-sample test shows that the two LFs differ over the range $i < 26$ at more
than 99\% confidence, but the strongest statement we feel is justified from
the present data is that the bright-end slope for the red clusters is somewhat
flatter than for the blue ones.  A better control field, and perhaps even better
selection of GCs versus field contamination, will be needed to carry this
discussion any further.

\subsection{Specific Frequencies}

Finally, we use the radial distributions we have derived
for NGC 3311 and NGC 3309 to estimate the total cluster populations
around each one, and thus their specific frequencies.  
We use the best-fit Sersic profile parameters given above,
and integrate them outward to large enough radius $r_{max}$ 
that the GC population totals converge.  
For NGC 3309,
the particular choice of $r_{max}$ is quite unimportant, since
its profile is steep and negligible numbers of GCs are added
for $r \gtrsim 3'$.  However, the much more extended profile for
NGC 3311 requires us to integrate outward to $r_{max} \simeq 9'$
or about 140 kpc, beyond which the Sersic model profile
adds negligible numbers of GCs.  This large radius is, however, quite consistent
with the existing surface photometry \citep{richter82,vasterberg91}
which shows that NGC 3311 extends detectably to $r \sim 150$ kpc.
With these parameters in mind, we find total GC populations
brighter than $i'=25.4$ equal to $N_{3311} = 7440$ and
$N_{3309} = 165$.  

Evaluating the actual population totals requires correcting for the
magnitude limit and the fraction of GCs fainter than that limit.
Assuming a Gaussian-like GCLF as discussed in the previous
section, with a turnover
point at $M_V \simeq -7.4 \pm 0.1$ and standard deviation 
$\sigma_V \simeq 1.5$, then $i'=25.4$ is
$0.25 \pm 0.1$ mag or $0.17 \sigma$ brighter
than the turnover point.  This part of the GCLF includes a fraction
$(0.45 \pm 0.04)$ of the total population.
We therefore estimate the total GC population for NGC 3311 as
$N_t \simeq 16500 \pm 2000$, where the main uncertainty is the 
necessary extrapolation to large radius.

For NGC 3309 we find $N_t = 364 \pm 210$;
here, the relatively large internal error is dominated instead by
the large uncertainty in the ratio 
$(\sigma_{e2}/\sigma_{e1}) = 0.080 \pm 0.046$ for the two-function fit
described above.   In this case our calculation may be a slight underestimate,
not because of what may be present at larger radius, but 
because of the uncertain {\sl inward} extrapolation into the central
$\sim 20''$. Its GCS core radius is both smaller and considerably more
uncertain (by a factor of two; see above), and our profile integration
may not have accounted for a significant part of the total.  
The HST photometry to be discussed in the next section
verifies that more GCs are present in its core.

Their specific frequencies
can be determined once we know their total luminosities.
\citet{vasterberg91} measure integrated magnitudes
$V_T = 11.04$ for NGC 3311 and 11.94 for NGC 3309
(we have reconstructed these numbers from their Table 5 and their
adopted distance and reddening).  
With our adopted $(m-M)_V = 33.88$,
their luminosities are $M_V^T(N3311) = -22.8$ and $M_V^T(N3309) = -21.9$.
The (global) specific frequency of the NGC 3311 GCS is then    
$S_N = N_t \cdot 10^{0.4(15-M_V^T)} = 12.5 \pm 1.5$, and for NGC 3309
$S_N = 0.63 \pm 0.36$.  It should be noted, however, that
these total magnitudes are extrapolations to large radius of
the $r^{1/4}$ profiles that were found to fit the inner regions,
and for NGC 3311 at least a part of the excess cD envelope light
would be missed.  Vasterberg et al.~estimate that it should
be made $\simeq 0.4$ mag brighter, which would give $M_V^T(N3311) = -23.2$.
If this adjustment is applied, its specific frequency would be lowered
to $S_N = 8.7 \pm 1.0$. However, it is not clear that the luminosity
should be increased
without also making a similar adjustment to the total cluster
population, since we might have missed
some of the GCs by using only the extrapolation from
the inner profile.  For the present time, and pending a more
quantitative measurement of the cD envelope light, we use
$S_N \simeq 12$, while $S_N \sim 9$ should be considered
an extreme lower limit.

By contrast, NGC 3309 has a remarkably low deduced cluster population,
even with the uncertainties about the core population.
Giant E galaxies in rich cluster environments with $S_N \lesssim 1$
are rare, but not unprecedented.  For example, \citet{baum95} found
$S_N < 1$ for one of the Coma cluster giants NGC 4881.
It is tempting to imagine that an object like NGC 3309 might have had
much of its original GCS removed by tidal stripping in passages through
the cluster core.  Regardless of its history, it is clear that today,
NGC 3311 completely dominates the GCS population
in the core Hydra field, in equal measure because of 
its higher total luminosity and its higher intrinsic $S_N$.
The wide ranges of $S_N$ that we find among these giant and structurally simple
E galaxies are still not fully understood and may require combinations of
formation conditions, later dynamical evolution in the potential field
of their cluster, and accounting for the presence or absence of hot halo
gas \citep[e.g.][]{harris01,mcl99}.

As a final test of the relative GC populations of the two galaxies,
we ran another solution for Eq.~\ref{fiteqn} where we imposed the {\sl assumption}
that the total number of GCs in each galaxy was in proportion to the
luminosity of the galaxy; that is, they have the same $S_N$.  
This requirement converts to 
$N_{GC}(3311) / N_{GC}(3309) \simeq 2.5$.
The best-fit solution gave $\chi_{\nu}^2(min) \simeq 2.8$, worse than our
optimum solutions described above ($\chi_{\nu}^2 = 2.0$) though not dramatically 
so.  The implication
is simply that the parameters for the NGC 3309 system are poorly constrained
within broad limits, a statement that is already reflected in the large 
uncertainty on its deduced best-fit cluster population.  

NGC 3311 clearly holds one of the very largest globular cluster systems
in the local universe, rivalling that of the classic high$-S_N$ prototype
M87 which has $N_t \simeq 13000$ and $S_N \simeq 13$ 
\citep{mcl94,tam06}.  The NGC 3311
GCS is spatially very extended, marked by a huge 
core radius $c = 94'' = 24$ kpc
or effective radius $r_e = 177'' = 45$ kpc also comparable to that
of the GCS in M87 \citep{mcl99}.

\section{WFPC2 Photometry}
\label{wfpc2}

To gain additional insight into the relative properties of the GCSs
around NGC 3309 and 3311, we employed imaging in $(V,I)$ from the
HST WFPC2 camera that was available in the HST Archive.
Two WFPC2 fields were taken as part of GO program 6554 (PI Brodie)
in Cycle 6, each with total exposures of 3700 sec in $F55W$ and
3800 sec in $F814W$.  In one field, the PC1 chip was centered on NGC 3311,
and in the second field, the PC1 chip was centered on NGC 3309.
Photometry of the NGC 3311-centered field was published in 
\citet{brodie00}, but to our knowledge, no measurements from the NGC 3309 imaging
have been published.

\begin{figure} 
\figurenum{9} 
\plotone{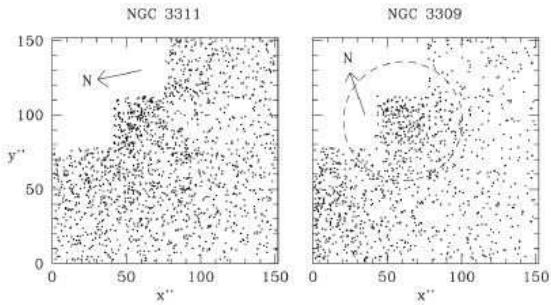} 
\caption{WFPC2 fields in the Hydra core.  The panel on the
left shows the data for the field in which the PC1 chip is
centered on NGC 3311, while the right panel shows the field
centered on NGC 3309.  All measured objects brighter than
$I = 25.0$ are plotted.  The $xy$ positions on the frames
are plotted in arcseconds.  Notice that the orientations of
the two fields are different:  for NGC 3311, North is
$109.8^o$ counterclockwise from vertical, and for NGC 3309,
North is $19.7^o$ counterclockwise from vertical.
Within the ``NGC 3309'' field, the center of NGC 3311 lies
to the lower left; notice the increase in numbers of measured
points there.
}
\label{wfpc2_xy}
\end{figure}

\begin{figure} 
\figurenum{10} 
\plotone{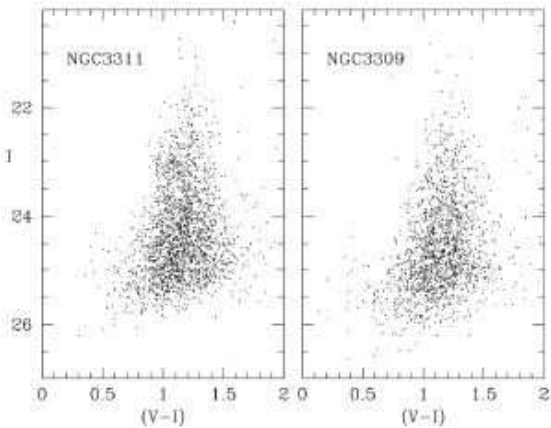} 
\caption{Photometry for the measured starlike objects in the two WFPC2 fields
shown in the previous figure.
}
\label{wfpc2_2cmd}
\end{figure}

We have carried out photometry on both of these WFPC2 fields with
{\sl daophot} in the same manner as described above.
A typical globular cluster with a half-light radius of $\sim 3$ pc
will, at the distance of Hydra, have an intrinsic FWHM $< 0\farcs03$,
making it entirely feasible to perform PSF-fitting photometry with
the WFPC2 resolution of $0\farcs1$.  For each of the four WFPC2 CCDs,
we registered and co-added the individual exposures downloaded from
the HST Archive, and performed the usual sequence of {\sl daophot}
FIND/PHOT/PSF/ALLSTAR.  Detected objects on the $V$ and $I$ frames
that matched locations within 2 pixels, and with ALLSTAR goodness-of-fit
$\chi_{V,I} < 2.0$, were kept for the final data list.
Calibrations followed the normal prescriptions
in \citet{holtzman95}, based on $0\farcs5$ aperture photometry of
the brightest objects.  The data for the four CCDs were then combined
to yield the results shown in Figures \ref{wfpc2_xy}
and \ref{wfpc2_2cmd}.  
The photometric limit near $I\simeq 25$ or $i' \simeq 25.3$ is 
about half a magnitude shallower than for our GMOS data.  

\begin{figure} 
\figurenum{11} 
\plotone{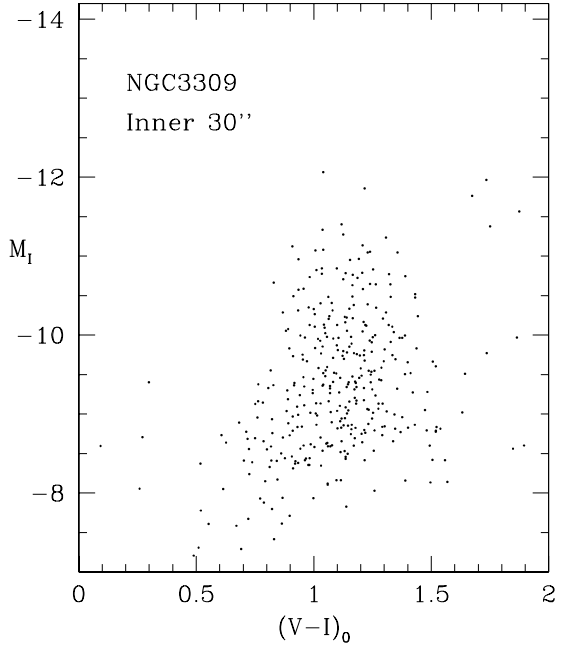} 
\caption{Photometry for measured objects within $30''$ of the center of NGC 3309.
}
\label{wfpc2_cmd_3309}
\end{figure}

Our PSF-based photometry for NGC 3311 has generated
a slightly tighter CMD than the work of \citet{brodie00} (who used
only aperture photometry), but no major differences.  Though the
$(V-I)$ color index is notoriously insensitive to metallicity
(with an intrinsic difference of only 0.2 mag between the red and blue
sequences), a clear trace of bimodality can be seen
\citep[also see][]{brodie00}.  Notably,
on the red side at $\langle V-I \rangle \simeq 1.2$, the upward
extension of the GC sequence into the high-mass UCD-like range
that is more evident in our more comprehensive GMOS photometry 
\citep{wehner07} can also be seen here.  In general, both CMDs
in Fig.~\ref{wfpc2_2cmd} resemble each other closely. 

In the $xy$ plots of Fig.~\ref{wfpc2_xy}, we see the ``NGC 3309
problem'' once again.  In the so-called NGC 3309 field, a population of GCs
is present on and near the PC1 chip centered on 3309; but the 
presence of the NGC 3311 GCS is extensive everywhere over the field
(note particularly the lower left region of the $xy$ plot where the increase in
GC number density toward the center of NGC 3311 is particularly obvious).
Even though the exposures were intended for study of the NGC 3309
system, the majority of detected objects in it actually belong
to NGC 3311.
From our quantitative double-profile solutions in the previous section, it can
easily be shown that for any location further than $40''$ from the
center of NGC 3309, the GCS of NGC 3309 contributes less than 10\%
of the GC counts.  This transition boundary beyond which
the NGC 3309 system becomes unimportant is indicated by the dashed
circle in Fig.~\ref{wfpc2_xy}.  The CMD for this inner zone
is shown in Figure \ref{wfpc2_cmd_3309}, where we have included
the $\simeq 360$ detected objects within an even smaller radius
of $30''$.  In this diagram, which gives us our best look at the GC population
of NGC 3309 alone, the high-luminosity end of the CMD is not as
well populated as in the NGC 3311 GCS (perhaps a simple result
of small-number statistics), but otherwise there are no striking
differences to note.

\begin{figure} 
\figurenum{12} 
\plotone{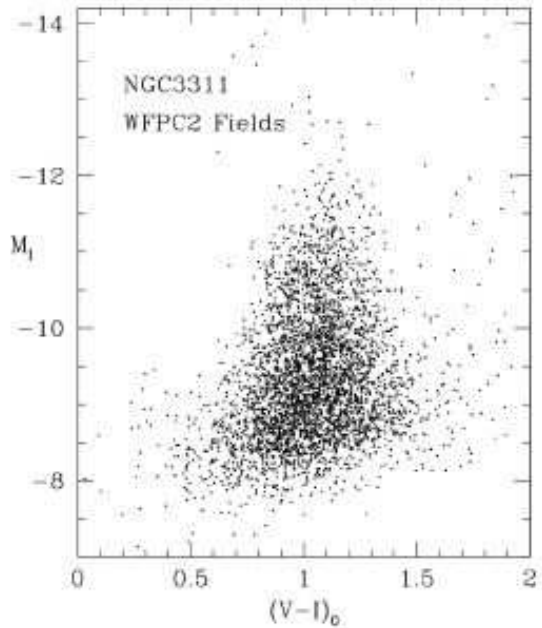} 
\caption{Color-magnitude data for both WFPC2 fields combined,
excluding any objects within $40''$ of the center of NGC 3309
(within the dashed circle in Fig.~\ref{wfpc2_xy}).
This CMD is completely dominated by GCs belonging to NGC 3311.
}
\label{wfpc2_cmd}
\end{figure}

In Figure \ref{wfpc2_cmd}, we have combined the photometry from both
WFPC2 fields, excluding only the points within the $40''$ zone around
NGC 3309.  This CMD, dominated by NGC 3311, will be analyzed below simultaneously with
the GMOS data, for bimodality characteristics of the metallicity
distribution.  

\section{The Metallicity Distribution and Analysis of the Bimodality}
\label{MDF}

A primary goal of our study was to investigate the multimodal
structure of the GC color and metallicity distribution, and in
particular to find out whether or not the blue GC sequence 
would reveal the same mass-metallicity relation as in other
large galaxies.  Having measured several other properties
of the system, including the fact that the GC population in our
target field is dominated by NGC 3311, we now discuss the color
distribution.

The histogram of colors $(g'-i')$ over any selected magnitude range
drawn from Fig.~\ref{cmd} indicates immediately that the color distribution is
broad and not well fit, for example, by a single Gaussian-type curve
\citep[also found by][from their WFPC2 photometry]{brodie00}.
A better description of the system is by conventional ``blue'' and ``red''
sequences whose color distributions partially overlap.  The bright-end
extension of the red sequence in particular 
reaches upward into the UCD luminosity range \citep{wehner07}.
We therefore experimented with bimodal fits in a variety of ways, to help define
the mean colors and dispersions of these two components.

To carry out quantitative fits to the color distribution, we employed the
statistical code RMIX \citep{macd07}, a library written with the statistics
programming language, R.  This package allows the user to fit data with
various functional shapes and to set constraints on them at a range of possible
levels.  RMIX does not restrict the fitting to unimodal or bimodal
forms; the number of components adopted
to fit the data is user-defined, as are the functional forms of the curve
(e.g. Gaussian, Poisson, binomial) and the starting values for the modes and
dispersions of each component (e.g. the peak and standard deviation, 
in the Gaussian case).  In addition, the 
dispersions can be constrained to be equal (homoscedastic) or
allowed to differ between components (heteroscedastic), as appropriate.  
The numerical values of the dispersions can also be set to previously determined
numbers, or can be determined by the fit, in any combination between the
components.\footnote{The complete code, available for a variety of platforms,
is publicly available from Peter MacDonald's website at 
http://www.math.mcmaster.ca/peter/mix/mix.html.  The same site gives links
to an extensive bibliography with further examples of its use.}

The KMM software \citep{mcl87,ash94} is a well known code frequently used
in the literature for bimodal fits to GCS color distributions
\citep[see, e.g.][among many such examples]{larsen01,mieske06}.
We have experimented with both KMM and RMIX in this study.  Both
give quite similar results for a given dataset, but we find
that the RMIX package offers more flexibility
in its choice of fitting functions as well as better flexibility in 
managing user-defined constraints on the fitting
parameters.  It is also more robust, successfully converging on multimodal
fits in small-sample cases where KMM would not run.  
We recommend RMIX as an attractive and statistically rigorous alternative.

\begin{figure} 
\figurenum{13} 
\plotone{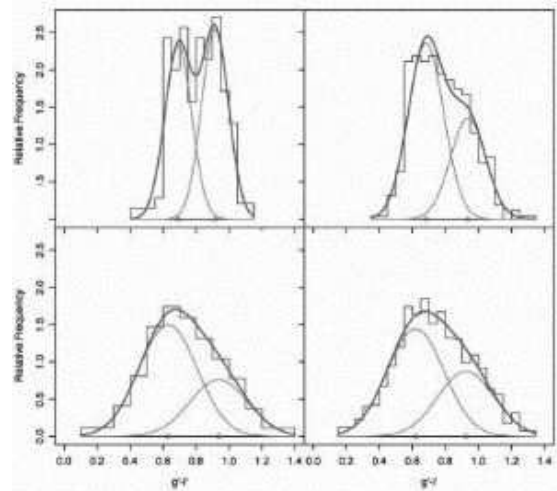} 
\caption{Color distribution for the globular cluster population
in NGC 3311, as a function of magnitude.  Number of GCs per
unit $(g'-i')_0$ interval is plotted for four different
luminosity bins as listed in Table 2.  The interval $i'=22-23$
is at upper left, $23-24$ at upper right, $24-25$ at lower
left, and $25-26$ at lower right.
In each panel, the bimodal Gaussian fit as
determined by the RMIX code is shown overplotted on the histogram
of the data.  The two individual Gaussians matching the metal-poor
and metal-rich components are shown by the red lines, while the
green envelope is the direct sum of the two components.
Small filled triangles on the x-axis of each plot mark the 
mean color (Gaussian peak position) of each component.
Note that the metal-poor component becomes progressively
redder at higher luminosity, while no significant change
occurs for the metal-rich component.
}
\label{mdf_4panel}
\end{figure}

Figure \ref{mdf_4panel} shows some of our results for four different 
magnitude subsets of our ($g'$,$i'$)
data.  For each magnitude bin shown, we fit a pair of Gaussian
curves to the $(g'-i')$ histogram. 
Reasonable initial guesses for the peaks and dispersions were
provided, and the component means and dispersions
were determined by the fit itself with the condition that
the dispersions of the two Gaussians were
required to be the same.  The results are summarized in Table \ref{rmixfits}.
Here, the first column gives the magnitude range; column 2 the deduced proportions
of the blue and red components; columns 3 and 4 the mean colors of the
blue and red sequences; column 5 the standard deviation of each sequence;
and the last column the goodness of fit.
Bimodal fits are strongly preferred over unimodal ones for the two
brighter bins. In the two faintest bins the overlap between the red
and blue components increases because of the combination of field contamination
and larger photometric uncertainty, and the distinction between bimodal and
unimodal fits becomes less clear.

\begin{deluxetable}{cccccc}
\tablecaption{RMIX fits to the Color Distribution \label{rmixfits}}
\tablewidth{0pt}
\tablehead{
\colhead{$i'$ Range} &  \colhead{f(blue,red)} &$\langle g'-i'\rangle_b $ 
& $\langle g'-i'\rangle_r$ & $\sigma_b, \sigma_r$ & $\chi_{\nu}^2$\\
} 
\startdata
$22-23$ & (0.48,0.52) & $0.850\pm 0.015$ &  $1.073 \pm 0.014$ & $0.082 \pm 0.008$ & 1.38 \\
$23-24$ & (0.64,0.36) & $0.840\pm 0.010$ &  $1.095 \pm 0.013$ & $0.108 \pm 0.006$ & 2.34 \\
$24-25$ & (0.62,0.38) & $0.778\pm 0.015$ &  $1.082 \pm 0.020$ & $0.172 \pm 0.008$ & 1.35 \\
$25-26$ & (0.72,0.28) & $0.705\pm 0.013$ &  $1.077 \pm 0.021$ & $0.186 \pm 0.007$ & 2.11 \\

\enddata

\end{deluxetable}

\begin{figure} 
\figurenum{14} 
\plotone{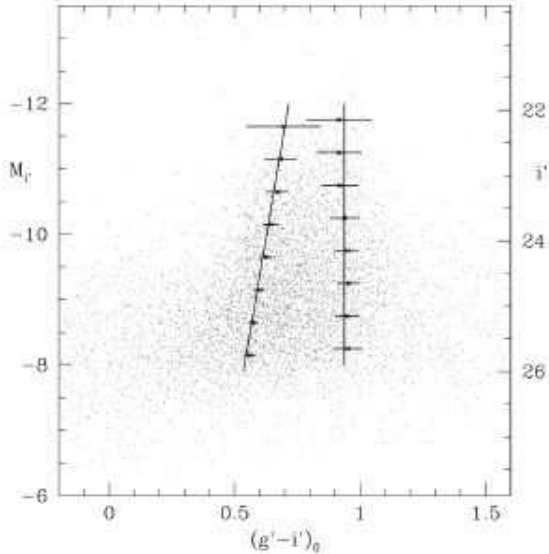} 
\caption{Color-magnitude diagram for the NGC 3311 globular
cluster system, plotted as absolute magnitude $M_{i'}$ versus
dereddened color.  Mean points for the blue and red
globular cluster sequences are overplotted, derived as
described in the text.  The error bars on each
point are the one-sigma uncertainties in $\langle g'-i' \rangle_0$.  
The best-fit mean lines determined by linear least squares
are drawn through each sequence.      For the blue sequence,
the slope is $\Delta(g'-i') / \Delta i' = 0.044 \pm 0.011$,
corresponding to a heavy-element abundance scaling with cluster
mass of $Z \sim M^{0.6}$.
The red GC sequence shows no net change in color with magnitude,
and is assumed for the purposes of this plot (see text and
previous figure) to be at a fixed color of $(g'-i')_0 = 0.94$.
}
\label{mmr}
\end{figure}

In Fig.~\ref{mdf_4panel}, one can see that the peak of the bluer component
shifts closer to the red component at 
brighter magnitudes, suggesting the presence of a
mass-metallicity trend for at least one of the modes.
To explore this further, we separated the GC population into blue and
red halves  by taking the intervals $0.3 < (g'-i')_0 < 0.8$ (blue)
and $0.8 < (g'-i')_0 < 1.2$ (red), and then
calculating the mean color of each mode in half-magnitude bins.
The combined results can be seen in Figure
\ref{mmr}.  Linear least-squares fits over the entire GC luminosity range
$22 < i' < 26$, weighted by $1/\sigma^2$ where the uncertainty
of each mean color is $\sigma$, give $\Delta(g'-i') / \Delta i' = 0.044 \pm 0.011$
for the blue sequence with a correlation coefficient
of the fit equal to 0.995, indicating that a MMR is present at high significance.   
For the red sequence, 
$\Delta(g'-i') / \Delta i' = 0.012 \pm 0.021$, indicating
no significant trend with magnitude. 
In Fig.~\ref{mmr}  we show it simply as
an average $(g'-i')_{0,red} = 0.94$.  

Field contamination at
magnitude levels $i' \lesssim 25$ is unimportant, so at these
relatively bright levels 
in the color-magnitude diagram the existence of an MMR seems
secure.  Fainter than this, however, contamination becomes 
progressively worse and is likely to affect these simple linear fits to 
uncertain degrees.  Thus for the fainter parts of the sequences
(roughly, $M_{i'} \gtrsim -9.5$) we cannot test with any confidence
whether or not the blue sequence really does continue 
to get ``bluer'' as it goes down.  
In \cite{har06}, we claimed that the blue sequence became approximately
vertical (that is, no MMR) for luminosity levels $M_I \gtrsim -9.5$. 
That claim was based on bimodal fits to the color distributions in finely spaced
magnitude bins for the eight BCGs studied there; however, in most of those
galaxies the issue of faint-end field contamination is also a concern
just as it is here.  For NGC 4594, \citet{spitler06} suggest that the 
blue-sequence MMR continues downward roughly linearly to fainter luminosities,
though their two faintest blue-sequence bins are not distinguishably
different in mean color, and their conclusion of the same MMR  at
all luminosities is in part forced by their assumption of a linear
fit.  To quote them directly, ``Indeed, whether the faintest bins
follow the trends established by the brighter points is indeterminate.''
For our NGC 3311 data, removing the two faintest bins, where we expect the contamination to be most significant, as well as the top bin, which may be affected by the presence of
Ultra Compact Dwarfs \citep{wehner07}, yields a 
similar fit of $\Delta(g'-i') / \Delta i' = 0.040 \pm 0.011$.  As such, the resulting
slope does not appear to be driven by contamination in the faintest magnitude bins.
Lastly, \citet{mieske06} use the combined GCS data for 79 Virgo 
galaxies to test for color trends by combining their target
galaxies into four groups by luminosity.  In their data, field 
contamination is much less of a concern because individual GCs were
carefully selected by a series of morphological and photometric
criteria.  For the highest-luminosity
galaxies (M87, M49, and M60; see their Figures 1 and 2) it can
reasonably be claimed that the blue-sequence MMR continues roughly
linearly over the entire luminosity range $-11 > M_{z'} > -7$.
For the other groups containing a total of 76 galaxies from giants
to dwarfs, the blue GC sequence clearly shows a color trend at
bright levels but could reasonably be claimed to
become nearly vertical for $M_{z'} \gtrsim -9$, corresponding roughly
to $M_{i'} \gtrsim -8.5$.  

At the present stage in this subject, we cannot yet say on the 
basis of the empirical evidence whether or not the blue-sequence MMR
persists at all luminosity levels.  More complete data allowing for
accurate removal of field contamination will be needed to trace it
out at the crucial faint levels.  All studies to date agree, however, that a
MMR exists for the blue clusters brighter than about $M_I \sim -9.5$,
and that this seems to be a nearly universal phenomenon.

Comparing our MMR for NGC 3311 with those found previously for other large
galaxies requires transformations between photometric systems, since
H06 used $(B-I)$, whereas \citet{mieske06} and \citet{strader06}
used $(g'-z')$.  Ideally, we would like to convert
$(g'-i')$ to either one of these other indices, but transformations
based directly on observations of standard globular clusters (the Milky Way
clusters, for example) do not yet exist.  For the purposes of
a preliminary comparison, we choose instead to transform the observed
blue-sequence slope $\Delta(g'-i')$ into a metallicity scale 
$\Delta$[Fe/H] through the use of models.  Representative 
calculations of integrated colors for SSP systems are
given by \citet{mar05} and \citet{gir04}.  For the range with
[m/H] $\lesssim -0.5$, these two sets of models agree well with
each other and
give $\Delta (g'-i') / \Delta {\rm [m/H]} = 0.18 \pm 0.03$ for
the metallicity regime we are concerned with, i.e.~[m/H] $\lesssim -1.2$.
Then, converting color index to metallicity, and magnitude to
luminosity and hence mass, we find that the MMR for the blue
sequence in NGC 3311 corresponds to a heavy-element abundance
scaling with cluster mass $Z \sim M^{0.6 \pm 0.2}$.
As we did in \citet{wehner07}, we assume that $(M/L)$ does not 
change systematically
with luminosity in order to convert $L$ to $M$ and thus derive
the scaling.
By comparison, H06 found a mean scaling $Z \sim M^{0.55}$ for their sample
of eight giant galaxies.  \citet{mieske06} found
$Z \sim M^{0.48 \pm 0.08}$ for their Virgo galaxy sample.
Our results for NGC 3311 are entirely 
consistent with the other studies within measurement uncertainties.

As an additional check of the $(g'-i')$ analysis, we perform the
same kinds of bimodal Gaussian fits to the WFPC2 data in $(V-I)$.
Here, the smaller size of the $(V-I)$ sample, and its lower sensitivity
to metallicity, prevent us from tracing out mean color trends with
magnitude with any confidence, so we use a color histogram defined
from only a single magnitude range $21.5 < I < 24.5$ over which
the field contamination and photometric scatter are minimized.  As before,
we exclude the region within $40''$ of NGC 3309.  

\begin{figure} 
\figurenum{15} 
\plotone{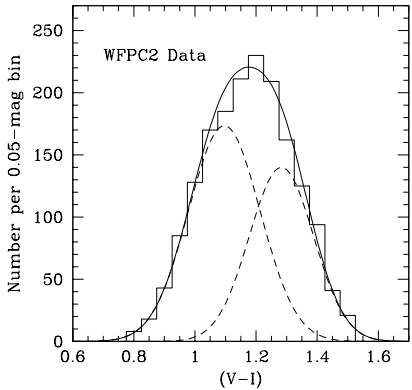} 
\caption{Color distribution for globular clusters in NGC 3311,
from the WFPC2 $(V-I)$ photometry discussed in the text.
No dereddening has been applied.  The double-Gaussian bimodal
fit to the data is shown, where the solid curve equals the
sum of the two individual components shown in the dashed lines.
The parameters for the two components are listed in the text.
}
\label{wfpc2_hist}
\end{figure}

The histogram of 1730 such objects is shown in Figure \ref{wfpc2_hist}.
We ran RMIX under similar assumptions as described above (two Gaussians,
with means, dispersions, and proportions determined by the fit).
For the blue, metal-poor component we find $\langle V-I \rangle = 1.10 \pm 0.07$,
$\sigma_{V-I} = 0.12 \pm 0.02$, and proportion 
of the total population $p(blue) = 0.58 \pm 0.39$.
For the red, metal-rich component, $\langle V-I \rangle = 1.28 \pm 0.07$,
$\sigma_{V-I} = 0.11 \pm 0.02$, and $p(red) = 0.42 \pm 0.39$, with an overall
$\chi_{\nu} = 1.22$.  The relative proportions of the two components
are quite uncertain -- again, a direct result of the insensitivity of
$V-I$ to metallicity and thus the heavy overlap between the two
components -- but agree nominally with the near-equal 
proportions we found from the $(g'-i')$ solutions (Fig.~\ref{mdf_4panel}).
For comparison purposes, we also tried a fit of a single Gaussian curve to the entire
data:  this yielded a best-fitting mean $\langle V-I\rangle = 1.175 \pm 0.004$ and
dispersion $\sigma_{V-I} = 0.146 \pm 0.003$, with goodness-of-fit
$\chi_{\nu} = 1.47$.  The bimodal fit performs better at matching
the data.

Although $(V-I)$ is less sensitive to metallicity, it has the advantage
over the SDSS indices that its conversion to [Fe/H] is more
well determined.  With $E_{V-I} = 0.11$, we have
$\langle V-I\rangle_0(blue) = 0.99 \pm 0.07$ and
$\langle V-I\rangle_0(red) = 1.17 \pm 0.07$.  Adopting a conversion
$(V-I)_0 = 0.16$ [Fe/H] $+ 1.15$ \citep{harris00,barmby00} based on
calibration from the Milky Way and M31 clusters then gives
[Fe/H] $\simeq -1.2$ for the blue sequence and $\simeq +0.2$ for
the red sequence.  These mean metallicities
should be taken only as rough estimates, since the
red-sequence mode in particular requires an extension of
the conversion equations 
into a high-metallicity regime that the calibrating GCs in M31
and the Milky Way do not adequately cover.  
However, the mean $(V-I)_0$ values themselves are close to   
what would be predicted from the established correlations with
host galaxy luminosity \citep{larsen01,strader04}, namely
\begin{eqnarray}
\langle V-I\rangle(blue) \, = \, -0.011 M_V^T \, + \, 0.71, \\
\langle V-I\rangle(red) \, = \, -0.018 M_V^T \, + \, 0.80 
\end{eqnarray}
For $M_V^T(NGC3311) = -22.8$ these relations predict
$\langle V-I\rangle_0 = 0.96$ and 1.19 for the blue and red modes
respectively, in good agreement with our observed mean colors. 
Supergiant ellipticals such as this one are the most logical places
to search for GCs of super-Solar metallicity, although higher quality
spectra than currently exist will be required to establish their
abundances accurately.

\section{Summary}
We have presented an extensive new photometric study of the large globular
cluster system around NGC 3311, the central cD galaxy in the Hydra cluster.
The relatively wide field and deep limiting magnitude of our data from
the Gemini/GMOS camera has allowed us to analyze the GCS with a sample
of several thousand globular clusters; our data reach to 
$i'(lim) \simeq 26.3$, just beyond the    
turnover point of the GC luminosity function.  Among our findings are
these:
\begin{itemize}
\item{} The classic metal-poor blue GC sequence
and the metal-richer red sequence are both present at their expected
colors, and  with nearly equal populations.  We show the color distributions
both from our GMOS $(g'-i')$ photometry and from HST WFPC2 archival data
in $(V-I)$.  
\item{} Bimodal fits to the
complete color distribution, subdivided by magnitude, confirm that the
blue sequence has the same correlation of progressively increasing
metallicity with GC mass that was previously found in many other
large galaxies; the correlation we find corresponds to a scaling of
GC metallicity with mass of $Z \sim M^{0.6}$.
\item{} By contrast, the red sequence shows no change of mean metallicity
with mass, though it does extend upward to much higher
than normal luminosity into the UCD-like range, strengthening the
potential connections between massive GCs and UCDs.
\item{} Confirming previous but much more preliminary investigations,
we find that the GCS in NGC 3311 falls definitively into the 
``high specific frequency'' category inhabited almost uniquely by
supergiant ellipticals (cD's or BCG's) at the centers of rich clusters.
NGC 3311 has an estimated total population of about 16,000 clusters
and a specific frequency $S_N \simeq 12$.
\item At the core of the Hydra cluster, another giant elliptical NGC 3309
is sitting projected just $100''$ from the cD NGC 3311.  We use our 
database to solve simultaneously for the spatial structure and total GC
populations of both galaxies at once.  The results show that NGC 3309
contributes only a few percent of the GC population over the entire field,
and has an unusually (though not unprecedentedly) low value
$S_N(N3309) = 0.6 \pm 0.4$. The central cD is completely dominant
in the central Hydra region.
\item{}  The GC luminosity function, which we
measure down to the ``turnover'' point at $M_I \simeq -8.4$, also
has a normal structure relative to other giant ellipticals.
\end{itemize}

Additional observations that could make clear improvements to the
understanding of the NGC 3311 system would include imaging with
a still wider field (to establish a true ``control'' field for
background measurement), and multicolor photometry with still more
sensitive color indices to improve the definition of the metallicity
distribution function.  The resources held by this enormous and
fascinating GC system are far from being exhausted.

\acknowledgements

The authors would like to thank the anonymous referee for their helpful comments in improving this paper.  This work is based on
observations obtained at the Gemini Observatory, which is operated by the
Association of Universities for Research in Astronomy, Inc., under a cooperative agreement
with the NSF on behalf of the Gemini partnership: the National Science Foundation (United
States), the Science and Technology Facilities Council (United Kingdom), the
National Research Council (Canada), CONICYT (Chile), the Australian Research Council
(Australia), CNPq (Brazil) and CONICET (Argentina).
EMHW, WEH, and KAW thank the Natural Sciences and Engineering Research Council of 
Canada for financial support.

\end{document}